\documentclass[twocolumn,aps,graphicx]{revtex4}%preprint,
\usepackage{graphicx}
\begin{document}

%\begin{CJK*}{GBK}{song}

\title{Quantum secure direct communication with quantum encryption based on pure entangled states
\footnote{Published in \emph{Chin. Phys.} \textbf{16}, 2149-2153
(2007)}}

\author{Xi-Han Li$^{a)b)c)}$,   Chun-Yan Li$^{a)b)c)}$, Fu-Guo  Deng$^{a)b)c)}$\footnote{E-mail addresses: fgdeng@bnu.edu.cn}, Ping
Zhou$^{a)b)c)}$,  Yu-Jie Liang$^{a)b)c)}$  and Hong-Yu
Zhou$^{a)b)c)}$}
\address{$^{a)}$ The Key Laboratory of Beam Technology and Material
Modification of Ministry of Education, Beijing Normal University,
Beijing 100875, China\\
$^{b)}$ Institute of Low Energy Nuclear Physics, and Department of
Material Science and Engineering, Beijing Normal University,
Beijing 100875,  China\\
$^{c)}$ Beijing Radiation Center, Beijing 100875, China}
\date{\today }

\begin{abstract}
We present a scheme for quantum secure direct communication with quantum encryption. The two authorized users
use repeatedly a sequence of the pure entangled pairs (quantum key) shared for encrypting and decrypting the
secret message carried by the traveling photons directly. For checking eavesdropping, the two parties perform
the single-photon measurements on some decoy particles before each round. This scheme has the advantage that the
pure entangled quantum signal source is feasible at present and any eavesdropper cannot steal the message.\\
\bigskip


\textbf{PACC:} 0155, 0367
\end{abstract}
\maketitle

Quantum mechanics provides some novel ways for processing and
transmission of quantum information. Quantum key distribution (QKD)
is considered to be the safest system for creating a private key
between two remote authorized users, say Alice and Bob, and may be
the most advanced application of quantum information. The noncloning
theorem forbids an eavesdropper, Eve to eavesdrop the quantum
communication freely. In 1984, Bennett and Brassard \cite{BB84}
proposed an original QKD protocol with nonorthogonal polarized
single photons. As an unknown quantum states cannot be eavesdropped
without leaving a trace in the outcomes obtained by the two parties,
the BB84 protocol is unconditionally secure \cite{BB84security}. In
1991, Ekert \cite{Ekert91} introduced another QKD scheme based on
the correlation of an Einstein-Podolsky-Rosen (EPR) pair, the
maximally entangled two-particle state, by using Bell inequality for
error rate analysis. Subsequently, Bennett, Brassard and Mermin
(BBM92) \cite{BBM92} simplified the process for eavesdropping check
in this scheme with two nonorthogonal measuring bases. Lo, Chau and
Ardehali \cite{ABC} presented a QKD model with two nonsymmetric
bases. Huang \emph{et al.} \cite{MBE} and Deng \emph{et al.}
\cite{CORE} designed two QKD models by using repeatedly a sequence
of private classical bits shared initially for improving their
efficiency for qubits or reducing the delay time in QKD with some
orthogonal states. To date, much attention has been focused on QKD
\cite{RMP,LongLiu,ABC,MBE,CORE,BidQKD,delay}.

Recently, quantum secure direct communication (QSDC), a novel
concept, was proposed and pursued by some groups
\cite{beige,pp,W,dengcp,two-step,QSDC,
caiqy1,caiqy2,LIXH,dengnetwork,GaoYan,Wangc,zhangzj,zhangs2006,song,gaocp,lijkps,wangj}.
Different from QKD whose goal is to generate a private key between
the two remote parties of communication, QSDC is used to communicate
the secret message directly without generating a key in advance and
then encrypting the message for its transmission in a classical
channel. According to the Deng-Long-Liu criterion
\cite{two-step,QSDC,dengnetwork}, on one hand, the sender Alice
should confirm whether the quantum channel is secure before she
encodes the message on the quantum states transmitted as the
messages cannot be discarded in QSDC
\cite{two-step,QSDC,dengnetwork}. Moreover, the message should be
read out by the receiver Bob directly
\cite{two-step,QSDC,dengnetwork}.  On the other hand, the security
of quantum communication is based on the error rate analysis with
the theories in statistics \cite{two-step,QSDC,dengnetwork}. In this
way, the quantum states in QSDC should be transmitted in a quantum
data block, same as those in Refs.
\cite{two-step,QSDC,dengnetwork,Wangc,LIXH,GaoYan,zhangzj}. It is
not a necessary condition in QKD as the error rate analysis is just
a postprocessing step, and the results transmitted do not include
the information about the secret message and can be abandoned
\cite{two-step,QSDC,dengnetwork}.

In 2002 Beige \emph{et al.} \cite{beige} proposed a QSDC scheme with
single-photon two-qubit states. In their scheme, one qubit is used
to check the security of the quantum channel after some results are
transmitted, and the other is used to carry  the message. The
message can be read out after a transmission of an additional
classical bit for each qubit \cite{pp}. Subsequently
Bostr$\ddot{o}$m and Felbingeer proposed a quasi-secure QSDC
protocol, called Ping-Pong QSDC protocol \cite{pp}, with an EPR
pair. In this scheme the eavesdropping  done by Eve can be hidden
with the loss of a practical quantum channel \cite{W} and the error
aroused by the noise of the quantum channel \cite{dengcp}. Moreover,
the two parties can accomplish the error rate analysis only after
they transmit an enough large set of message, and some message  may
leak to Eve in a noise channel. The two-step QSDC scheme
\cite{two-step} and the scheme with single photons \cite{QSDC}
proposed by Deng \emph{et al.} can be made secure even in a noisy
channel with quantum privacy amplification \cite{deutschqpa,QPA} as
the states are transmitted in a quantum date block and the
confirmation of the security of the quantum channel has be achieved
before the sender encodes her message on the states. These ideas are
borrowed by Wang \emph{et al.} \cite{Wangc} to design the QSDC
models as well.

In this paper, we will present a scheme for quantum secure direct
communication with quantum encryption. This protocol uses a
controlled-not (CNot) gate to encode and decode the secret message.
The two parties first share privately a sequence of two-photon pure
entangled states which are feasible with the present technologies,
and then use the states
 as their private quantum key which is reusable with a
eavesdropping check before each round. The receiver can read out
directly the message and each photon transmitted between the parties
can carry one bit of message securely in principle.

Generally speaking, many QKD and QSDC schemes choose EPR pairs as
quantum information carriers to transmit a private key or a secret
message. The reason is that the two photons in the four Bell states
have good correlations in both the measuring bases $Z=\{\vert 0
\rangle,\vert 1 \rangle\}$ and $X=\{\vert \pm x
\rangle=\frac{1}{\sqrt{2}}(\vert 0\rangle \pm \vert 1\rangle)\}$.
\begin{eqnarray}
\vert \phi^+ \rangle_{AB}&=& \frac{1}{\sqrt{2}}(\vert 0 \rangle_A
\vert 0 \rangle_B + \vert 1 \rangle_A \vert 1 \rangle_B)\nonumber\\
&=& \frac{1}{\sqrt{2}}(\vert +x \rangle_A \vert +x \rangle_B + \vert
-x \rangle_A \vert -x \rangle_B)\\
\vert \phi^- \rangle_{AB}&=& \frac{1}{\sqrt{2}}(\vert 0 \rangle_A
\vert 0 \rangle_B - \vert 1 \rangle_A \vert 1 \rangle_B)\nonumber\\
&=& \frac{1}{\sqrt{2}}(\vert +x \rangle_A \vert -x \rangle_B + \vert
-x \rangle_A \vert +x \rangle_B)\\
\vert \psi^+ \rangle_{AB}&=& \frac{1}{\sqrt{2}}(\vert 0 \rangle_A
\vert 1 \rangle_B + \vert 1 \rangle_A \vert 0 \rangle_B)\nonumber\\
&=& \frac{1}{\sqrt{2}}(\vert +x \rangle_A \vert +x \rangle_B - \vert
-x \rangle_A \vert -x \rangle_B)\\
\vert \psi^- \rangle_{AB}&=& \frac{1}{\sqrt{2}}(\vert 0 \rangle_A
\vert 1 \rangle_B - \vert 1 \rangle_A \vert 1 \rangle_B)\nonumber\\
&=& \frac{1}{\sqrt{2}}(\vert +x \rangle_A \vert -x \rangle_B - \vert
-x \rangle_A \vert +x \rangle_B)
\end{eqnarray}
This feature is very useful in the eavesdropping check process.
However, it is difficult to produce the maximally two-particle
entangled states with the present technologies. Contrarily, in
experiments the two photons prepared are usually in the pure
entangled state, such as $\vert \Psi \rangle_{AB}=a \vert 0
\rangle_A \vert 0 \rangle_B + b \vert 1 \rangle_A \vert 1 \rangle_B$
($\vert a\vert^2 + \vert b \vert^2 = 1$). The two photons are always
correlated in the basis $Z$, but not in the basis $X$, as
\begin{eqnarray}
\vert \Psi \rangle_{AB}&=& a\vert 0 \rangle_A \vert 0 \rangle_B + b
\vert 1 \rangle_A \vert 1 \rangle_B\nonumber\\
&=&\frac{1}{2}[(a+b)(\vert +x \rangle_A \vert +x \rangle_B +\vert -x
\rangle_A \vert -x \rangle_B)\nonumber\\
&+&(a-b)(\vert +x \rangle_A \vert -x \rangle_B +\vert -x \rangle_A
\vert +x \rangle_B)].
\end{eqnarray}
This means that the security of the quantum communication with pure
entangled states is lower than that with Bell states if we use the
two bases $X$ and $Z$ to check the security. On the other hand,
those pure entangled quantum sources are more convenient in
practical applications. In this QSDC scheme, we use some decoy
photons to check the eavesdropping for overcoming the flaw of pure
entangled states.

Now, let us describe our QSDC scheme in detail as follows. It
includes four steps.

(S1) The two authorized users (Alice and Bob) share a sequence of
two-particle pure entangled states securely as their private quantum
key.

For this task, Bob first prepares $n$ two-photon pairs randomly in
one of the two pure entangled states $\vert \Psi \rangle_{AB}$ and
$\vert \Phi \rangle_{AB}$. Here $\vert \Phi \rangle_{AB}=b \vert 0
\rangle_A \vert 0 \rangle_B + a \vert 1 \rangle_A \vert 1 \rangle_B$
which can be obtained by flipping the bit value of the two photons
in the state $\vert \Psi \rangle_{AB}$; i.e., $\vert \Phi
\rangle_{AB}=\sigma_x^A\otimes \sigma_x^B \vert \Psi \rangle_{AB}$
(here $\sigma_x$ is a Pauli operator). Bob picks up photon $B$ in
each pair for forming the sequence $S_B:$ $[B_1,B_2,\cdots,B_n]$.
The other sequence $S_A$ is made up of particles $A_i(i=1,2...n)$.
He keeps the sequence $S_B$ at home and sends the sequence $S_A$ to
Alice. For checking eavesdropping, Bob inserts some decoy photons
$S_{de}$, which are randomly in one of the four states $\{\vert
0\rangle, \vert 1 \rangle, \vert +x \rangle, \vert -x \rangle\}$,
into the sequence $S_A$. He can get a decoy photon by measuring one
photon in a two-photon pair $\vert \Psi \rangle_{AB}$ with the basis
$Z$  and operating on the other photon with $\sigma_x$ or a Hadamard
(H) operation. In a word, it is unnecessary for the users to have an
ideal single-photon source in this scheme.

After Alice announces the the receipt of the sequence $S_A$, Bob
tells her the positions and the states of the decoy photons. Alice
measures the decoy photons with the suitable bases and analyzes the
error rate of those outcomes with Bob. If the error rate is very
low, they can obtain a sequence of quantum key privately and
continue to the next step; otherwise, they discard the transmission
and repeat quantum communication from the beginning.

(S2) Alice and Bob use their private quantum key to encrypt and
decrypt the secret message directly.

For QSDC, Alice prepares a sequence of traveling particles
$\gamma_{i}$ which are in one of the two states $\{\vert 0 \rangle,
\vert 1 \rangle\}$ according to the bit value of her secret message
is 0 or 1, respectively. We call it the traveling particle sequence
$S_T$. For checking eavesdropping, similar to Refs.
\cite{two-step,QSDC}, Alice needs to add a small trick in the
sequence $S_T$ before she sends it to the quantum channel. That is,
he inserts some decoy photons, say $S_D$ which are randomly in the
four states $\{\vert 0\rangle, \vert 1\rangle, \vert +x \rangle,
\vert -x\rangle\}$ and distributed randomly in the sequence $S_T$.
Alice uses the quantum key, the pure entangled pairs shared $\{\vert
\Psi \rangle_{AB}, \vert \Phi \rangle_{AB}\}$ to encrypt the
traveling particles in the sequence $S_T$ except for the decoy
photons. That is, Alice performs a CNot operation on the particles
$A_i$ and $\gamma_i$ ($i=1,2,\ldots, n$) by using the particle $A_i$
as the control qubit. Then Alice sends all the traveling particles
to Bob.

After receiving the sequence $S_T$, Bob asks Alice to tell him the
positions and the states of the decoy photons in the sequence $S_T$,
and then measures them with the same bases  as those Alice chose for
preparing them. For the particles $B_i$ and $\gamma_i$, Bob takes a
CNot operation on them with the particle $B_i$ as the control qubit,
similar to Alice, and then he measures the particles $\gamma_i$ with
the basis $Z$ and records the outcomes of the measurements.

(S3) Alice and Bob check the security of their transmission. They
analyze the error rate of the decoy photons. If the error rate is
very low, Bob can trust this transmission, and read out the message
directly. Otherwise they have to abandon the results and repeat the
procedures from the beginning.

(S4) If the quantum communication succeeds, Alice and Bob repeat
their communication from step 2. That is, they use repeatedly the
pure entangled pairs as their quantum key and transmit the secret
message again in the next round.

In an ideal condition, this QSDC scheme is secure for any
intercepting-resending attack strategies if the quantum key is
private because it is equivalent to a quantum one-time-pad
crypto-system \cite{RMP,QSDC}. The encryption on a traveling
particle $\gamma_i$ with a CNot operation makes it entangle with a
quantum system in the quantum key. For an eavesdropper, any quantum
system in the quantum key is randomly in one of the two states
$\{\vert \Psi\rangle_{AB}=a\vert 0\rangle_A\vert 0\rangle_B + b\vert
1\rangle_A\vert 1\rangle_B, \vert \Phi\rangle_{AB}=b\vert
0\rangle_A\vert 0\rangle_B + a\vert 1\rangle_A\vert 1\rangle_B\}$.
After the CNot operation done by Alice, the three particles $A_i$,
$B_i$, and $\gamma_i$ are in the GHZ-class state
$\left\vert\Psi\right\rangle_{s1}=a \vert 00\gamma_i\rangle + b\vert
11 \overline{\gamma_i} \rangle)_{A_iB_i\gamma_i}$ or
$\left\vert\Phi\right\rangle_{s1}=b \vert 00\gamma_i\rangle + a\vert
11 \overline{\gamma_i} \rangle)_{A_iB_i\gamma_i}$ randomly. The
reduced density matrix of the traveling particle $\gamma_i$ is
$\rho_{\gamma_i}=\frac{1}{2}\left(\begin{array}{cc}
1 &  0\\
0 & 1
\end{array}
\right)$. That is, the traveling particle is in one of the two
eigenvectors of any measuring operator with the same probability
50\% for any eavesdropper, and none can get a useful information
about the state of the traveling particle $\gamma_i$ as the effect
of his eavesdropping on the traveling particle is as same as that by
guessing its outcome randomly.

If Eve wants to steal the message carried by the traveling
particles, she should eavesdrop the quantum key by capturing the
traveling particles first, and then steal the information of the
traveling particles transmitted in the next round. However, her
action will leave a track in the eavesdropping check. The reason is
that  the procedure of checking eavesdropping done by the two
authorized users in this scheme, in essence, is the same as that in
the BB84 QKD protocol \cite{BB84} for any eavesdropper. That is, the
decoy photons are produced by choosing randomly one of the two bases
$Z$ and $X$, and are inserted into the traveling sequence $S_T$
randomly. Also, Bob measures them with the same bases as those
chosen by Alice after he receives the traveling sequence and obtains
the information about their bases of the decoy photons. For any
eavesdropper, the bases chosen by Bob are random even though they
are announced in public finally as no eavesdropper has the access to
the traveling particles after Bob receives them. Any eavesdropping
will inevitably disturb the states of the decoy photons and be
detected by the two authorized users if an eavesdropper monitors the
quantum line, same as that in the BB84 protocol \cite{BB84}.

In a practical channel, noises will affect the entanglement of the
quantum key. In this time, the users should exploit entanglement
purification \cite{purification} to keep the entanglement in the
quantum key, and do quantum privacy amplification \cite{deutschqpa}
on them as well. However, the two users need not to purify their
states to Bell states, just the pure entangled states $\vert \Psi
\rangle_{AB}=a \vert 0 \rangle_A \vert 0 \rangle_B + b \vert 1
\rangle_A \vert 1 \rangle_B$ ($\vert \Phi \rangle_{AB}=b \vert 0
\rangle_A \vert 0 \rangle_B + a \vert 1 \rangle_A \vert 1
\rangle_B$) or $\vert \Psi' \rangle_{AB}=a' \vert 0 \rangle_A \vert
0 \rangle_B + b' \vert 1 \rangle_A \vert 1 \rangle_B$ ($\vert \Phi'
\rangle_{AB}=b' \vert 0 \rangle_A \vert 0 \rangle_B + a' \vert 1
\rangle_A \vert 1 \rangle_B$) in this scheme. Here $\vert a'\vert^2
+ \vert b'\vert ^2 =1$. As the quantum key is just used to encrypt
and decrypt the secret message, it is unnecessary for the users to
keep the same states as those they used in last time, just the
correlation of each pair, which will increase the efficiency of the
entanglement purification process largely. Certainly, on the one
hand, the users should do error correction on their results in
practical applications, same as the two-step protocol
\cite{two-step}. On the other hand, this QSDC scheme can only used
to distribute a private key if the loss of the quantum line is
unreasonably large.

In summary, quantum secure direct communication can be done with
quantum encryption by means that the two authorized users first
share privately a sequence of pure entangled pairs and use them
repeatedly as the quantum key for encrypting the traveling particles
which are in the eigenvectors of the basis $Z$. The receiver can
read out the message directly and each particle can carry one bit of
the message securely. For checking eavesdropping, the sender adds a
small trick in the traveling particles which can forbid Eve to
eavesdrop the quantum channel freely. The most important advantage
of this protocol is that the pure entangled quantum signal source is
feasible at present. As the photons used for security checking is
not the quantum key shared, the quantum key is reusable without
abatement in principle.

This work was  supported  by the National Natural Science Foundation
of China (Grant Nos. 10604008 and 10435020) and the Beijing
Education Committee (Grant No XK100270454).

%\end{CJK*}
\end{document}